\documentclass[12pt]{article}

\usepackage{mathrsfs,amsfonts,amssymb,amsmath,bm,graphicx}
\usepackage{fullpage,authblk,fullwidth}
\usepackage{epsfig,float,afterpage}

\renewcommand\a{\alpha}
\renewcommand\b{\beta}

\newcommand\del{\delta}

\renewcommand\k{\kappa}

\newcommand\m{\mu}

\renewcommand\t{\tau}

\renewcommand\j{\psi}

\newcommand\J{\Psi}

\renewcommand\d{\partial}

\newcommand{\lan}{\langle}
\newcommand{\ran}{\rangle}

\begin{document}

\title{Pancharatnam-Zak phase}

\author[1,2]{Vivek M. Vyas\thanks{physics.vivek@gmail.com}}

\author[1]{Dibyendu Roy} 

\affil[1]{Raman Research Institute, C. V. Raman Avenue, Sadashivnagar, Bangalore 560080, India}

\affil[2]{Indian Institute of Information Technology Vadodara, Government Engineering College, Sector 28, Gandhinagar 382028, India}

%providecommand{\keywords}[1]{\textbf{\textit{Index terms---}} #1}
\date{}
\maketitle

\begin{abstract}
	Three decades ago, in a celebrated work,  Zak found an expression for the geometric phase acquired by an electron in a one-dimensional periodic lattice as it traverses the Bloch band. Such a geometric phase is useful in characterizing the topological properties and the electric polarization of the periodic system. Unfortunately Zak's expression suffers from two flaws: its value depends upon the choice of origin of the unit cell, and is gauge dependent. Here we explain that these flaws in Zak's expression arise from the assumption that the electron's adiabatic motion is cyclic in the sense of recurrence of the density matrix in course of time evolution. We find through a careful investigation that the system displays cyclicity in a generalized sense wherein the physical observables return in the course of evolution. This notion of generalized cyclicity paves the way for a correct and consistent expression for the geometric phase in this system, christened as Pancharatnam-Zak phase. Pancharatnam-Zak geometric phase does not suffer from the flaws inherent in Zak's expression, and correctly classifies the Bloch bands of the lattice. A natural filled band extension of the Pancharatnam-Zak phase is also constructed and studied. 
\end{abstract}

\section{Introduction}

Recent years have witnessed a rapid growth in the application of the abstract concepts of topology in physics, particularly in condensed matter physics~\cite{bernevig2013topological, asboth2016short, vanderbilt2018}. It has led to the discovery of a new type of material, for example, topological insulators and the detection of hitherto unobserved Majorana and Weyl fermions as emergent quasiparticles in low-energy condensed matter systems. The \emph{geometric phase} is one such well-known example in physics where topological concepts enter in an essential way. The geometric phase was first anticipated by Pancharatnam~\cite{pancha1956,berry1987} and came to prominence due to Berry~\cite{berry1984}. The geometric phase provides insights into the curvature of the underlying Hilbert space of the quantum states. In many cases, it is found to act as a topological index, shedding light on the topological properties of the system~\cite{shapere1989}.

In landmark papers, Thouless {\it et al.} \cite{thouless1982,niut} showed that the electrical conductivity in a quantum Hall system could be understood in terms of a topological invariant, known as the Chern number. It was soon realized that the Chern number was closely related to the geometric phase~\cite{simon1983holonomy} acquired by the electron as it moved through the energy band. In the last three decades, the notion of geometric phase has been used to understand and classify the properties of several condensed matter systems ~\cite{shapere1989,haldane1988model,haldane2004berry}. The geometric phase plays a central role in the current understanding of materials like topological insulators and superconductors. Its value is inevitably found to govern the charge transport property in such systems~\cite{bernevig2013topological, asboth2016short, vanderbilt2018, thouless1983quantization, kane2010}.

In this paper, we provide a correct and consistent expression of the geometric phase acquired by an electron in a one-dimensional (1D) periodic lattice, as it executes a circuit over the Bloch band, which we call the \emph{Pancharatnam-Zak phase}. The earliest attempt in evaluating this geometric phase was due to Zak~\cite{zak1989}, and the expression obtained in that work is popularly called the \emph{Zak phase}. Later, King-Smith and Vanderbilt~\cite{king1993theory} showed that the difference of the Zak phase between two different configurations manifests in the treatment of quantized particle transport in a 1D insulator, a phenomenon discovered earlier by Thouless~\cite{thouless1983quantization}. Subsequently, the modern understanding of the change in electric polarizations in dielectric materials was formulated in terms of such a difference of Zak phase~\cite{vanderbilt2018, king1993theory, resta1994macroscopic, resta2000manifestations, niu}. The concept of Zak phase has been further applied to study the dynamics of strongly coupled LC circuits~\cite{tal2018} and waveguide lattices~\cite{longhi2013}. It has also been employed to classify the edge states in planar honeycomb lattice systems~\cite{delplace,kameda2019}.

% The Zak phase is also used for characterizing topological edge states of 1D photonic lattice for the experimental demonstration of lasing in topological edge states~\cite{Jean2017} and for isolating topological defects from trivial defects~\cite{Blanco2016}.   

Geometric phases have been observed and explored in several experiments in diverse areas of physics~\cite{shapere1989,bhandari1988}. The value of the geometric phase, a physically measurable quantity, can not depend upon choosing the origin of co-ordinates or gauge employed in evaluating it. There is a freedom to select the unit cell's origin in periodic lattice systems as shown later in Fig. \ref{schlattice}. All the physical observables are insensitive to such freedom of defining the unit cell, as it must be~\cite{ashcroftbook}. Nevertheless, it is a well-acknowledged fact in the literature that the Zak phase is a gauge-dependent object, and its value depends upon the choice of the 
origin of the unit cell ~\cite{vanderbilt2018, zak1989, resta2000manifestations, niu, dalibard, atala2013, moore2017}. As a result, it can attain any desired value by a suitable choice of the origin of the unit cell or an appropriate gauge choice for the Bloch states. To circumvent such an ambiguity of the Zak phase, a certain preferred choice of gauge and origin of the unit cell has been employed~\cite{vanderbilt2018,dalibard,atala2013,moore2017}. These observations only show that the Zak phase can not be a proper geometric object, let alone be physically observable. 

We must mention that the Zak phase difference between two different configurations/states of a system is independent of gauge and choice of unit cell. Such a difference of the Zak phase was experimentally observed by Atala \emph{et al.} \cite{atala2013}, and manifests in the change of electric polarizations \cite{vanderbilt2018,king1993theory}. Our main point in this paper is that the geometric phase itself is a well defined and measurable quantity, and it is correctly captured in the Pancharatnam-Zak phase for a 1D periodic lattice. %it does not rectify the unacceptable flaws inherent to the construction of the Zak phase. 

Motivated by the classic work of Berry~\cite{berry1984}, Zak, in his derivation, assumed that the adiabatic motion of an electron in a band was cyclic so that the recurrence of the initial state of the system (modulo an overall phase factor) happens over time evolution. However, we find 
that the underlying system displays cyclicity under time evolution in a 
generalized sense wherein the observables rather than the initial state or the density matrix return in the course of evolution.

Following the notion of generalized cyclicity, we here carefully consider the concept of the geometric phase in its generality and find the geometric phase gained by (a) a single electron and (b) by electrons of a filled band of the lattice when influenced by a weak electromagnetic field. The geometric phase in the single electron case - Pancharatnam-Zak phase, possesses the essential invariances under gauge transformation and unit cell reparametrization. The underlying geometrical and topological properties of the system are uncovered in the process. The Pancharatnam-Zak phase is found to act as a topological index for systems with inversion symmetry; it is either equal to $0$ or $\pi$ in the topologically trivial or non-trivial state, respectively. The geometrical phase for the filled band case is properly formulated, and its physical implications are discussed. In a later study, a generalization of the Pancharatnam-Zak phase is shown to be quite useful in capturing the topological phases of an undriven and periodically driven non-Hermitian Su-Schrieffer-Heeger (SSH) model~\cite{vmv2020}.

The paper is organized as follows. In section (\ref{panch}), the geometric phase concept, as defined in its generality, is briefly reviewed. Subsequently, the problem of a charged particle in a 1D periodic lattice, subjected to a weak electromagnetic field, is formulated, and its kinematic aspects are studied in section (\ref{gpsec}), bringing out the underlying mathematical structure. In section (\ref{gppsec}), the adiabatic quantum dynamics of such a motion discussed in (a) single-particle case and (b) many-particle filled-band case; and the manifestation of the geometric Pancharatnam-Zak phase in both the cases is found. An explicit calculation of the single-particle Pancharatnam-Zak phase for the SSH model is provided in section (\ref{eexam}), followed by the discussion in section (\ref{dis}). We further add two appendices for some details on the notion of geodesic and mathematical structure behind the Pancharatnam-Zak phase.

\section{Geometric phase via Pancharatnam route \label{panch}}

It has been long known that the notion of geometric phase $\gamma_{g}$ can be best understood following the work of Pancharatnam \cite{pancha1956,samuel1988,mukunda1993}. Pancharatnam's definition of the geometric phase $\gamma_{g}$ is given as an argument of a cyclic expression:
\begin{align} 
	\label{gpdef}
	\gamma_{g} = \text{Arg} \: \left( \lan \j_{0} | \j_{M} \ran \lan \j_{M} | \j_{M-1} \ran \cdots \lan \j_{2} | \j_{1} \ran \lan \j_{1} | \j_{0} \ran \right).
\end{align}
It is well defined for any given 
ordered set of vectors $| \j_{j} \ran$ for $j = 0,1, 2,\cdots, M$ (e.g., cell-periodic Bloch states or photon polarization states), 
provided only that the quantity in square brackets does not vanish. 
This definition shows that the geometric phase is a collective property
of an ordered set of vectors  $| \j_{j} \ran$ for $j = 0,1, 2,\cdots, M$. The definition does not rely on any dynamical aspect of the underlying system,
such as the Hamiltonian. Hence, one says that the notion of geometric phase
is kinematic in nature.

Clearly, $\gamma_{g}$ can not be altered by any redefinition of states $|\j_{j} \ran$:
\begin{align}
	|\j_{j} \ran \rightarrow e^{i \theta_{j}}|\j_{j} \ran,
\end{align} 
where $\theta_{j}$s are independent arbitrary real numbers. This property of (local) gauge invariance is a clear demonstration of the geometric nature of this phase. Evidently the geometric phase also remains invariant under unitary operations of the type:
\begin{align}
	|\j_{j} \ran \rightarrow \hat{U}|\j_{j} \ran,
\end{align}
which is the statement of basis independence of $\gamma_{g}$.  

If the set of states $|\j_{j} \ran \equiv |\j(s_{j}) \ran$ describes some quantum system at times $s_{j} = j \delta$ ($\delta$ is an infinitesimal time interval), then $\gamma_{g}$ is  
the geometric phase acquired by the system in course of evolution from time $s_0$ to $s_M$. In the continuum limit, $\gamma_{g}$ takes a familiar form:
\begin{align} \label{gpcont}
	\gamma_{g}(t) = \text{Arg} \lan \j(0) | \j(t) \ran + i \int_{0}^{t} ds \: \lan \j(s)| \partial_{s} | \j(s) \ran,
\end{align}
where $t = s_M$. This expression for the geometric phase and its proper generalization encompassing the case of non-unitary evolution were obtained long back using a manifestly geometric route~\cite{samuel1988}. The existence of such a geometric phase is well established through several experiments~\cite{berry1987, shapere1989,bhandari1988, chyba1988, martin2012}. 

%It is clear that the geometric phase $\gamma_{g}$ so defined by (\ref{gpcont}) is invariant under the \emph{local gauge transformation} $| \j(s) \ran \rightarrow e^{i \Lambda (s)} | \j(s) \ran$ (where $\Lambda(s)$ is some arbitrary function of $s$). It also possesses another invariance under the \emph{static unitary operations}. Let the state $| \bar{\j} \ran$ obtained from $| \j \ran$ as: $| \bar{\j} (s)\ran = \hat{T} | {\j} (s)\ran$, using a time independent unitary operator $\hat{T}$. It immediately follows that $\lan \bar{\j} (0) | \bar{\j} (t) \ran = \lan {\j} (0) | {\j} (t) \ran$, and $\lan \bar{\j} (s)| \d_s | \bar{\j} (s) \ran = \lan {\j} (s)| \d_s | {\j} (s) \ran$, which shows the invariance of $\gamma_{g}$ under action of $\hat{T}$. Since the coordinate translation operator $\hat{T}_{x}(\epsilon) = e^{- \frac{i \epsilon}{\hbar} \hat{p}}$ and momentum translation operator $\hat{T}_{p}(\epsilon) = e^{- \frac{i \epsilon}{\hbar} \hat{x}}$ are both time independent unitary operations, it thus follows that the geometric phase $\gamma_{g}$ is {invariant} under coordinate and momentum translations.

The geometric phase also possesses an important property, that of reparameterisation invariance. If a real parameter $r(s)$, an increasing function of time $s$, relabels the states $| \j(s) \ran$, so that $| \j (s) \ran \equiv | \varphi (r) \ran$, then one immediately sees that the geometric phase is invariant: 
\begin{align} \nonumber
	\text{Arg} \lan \j(0) | \j(t) \ran + i \int_{0}^{t}ds \: \lan \j(s) |\partial_{s}| \j( s) \ran
	= \text{Arg} \lan \varphi (r(0)) | \varphi (r(t)) \ran + i \int_{r(0)}^{r(t)}  dr \: \lan \varphi (r) | \partial_{r}| \varphi (r) \ran.
\end{align}

It is a well known fact, that two unit normalised states $|A\ran$ and $e^{i \lambda}|A\ran$ (for some arbitrary real $\lambda$), which differ by a phase, actually depict the same physical state of the quantum system \cite{messiah1999}. Often it is beneficial to work with density matrix $|A\ran \lan A |$ to describe the system since it is immune to such phase ambiguities. It is well known that there exist a notion of distance in the space of density matrices between any two density matrices \cite{samuel1988,mukunda1993}, which is summarised in Appendix~A for the benefit of the reader. In fact a well known result of Ref. \cite{samuel1988}, dictates that there exists a unique shortest curve - \emph{a geodesic}  connecting any two non-orthogonal states. Invoking this treatment, one finds that the geodesic curve $|\j'(s)\ran \lan \j'(s)|$ (for $0 \leq s \leq \Lambda$) connecting $|\j(t)\ran \lan \j(t)|$ and $|\j(0)\ran \lan \j(0)|$ is constructible from the states:
\begin{align} \label{geodesic} 
	|\j'(s)\ran = \frac{e^{i \theta s /\Lambda}}{\sin \Lambda} \left( \sin (\Lambda - s) |\j(t)\ran + e^{-i\theta} \sin (s) |\j(0)\ran \right).
\end{align}
Importantly the phase of the overlap $\theta =-\text{Arg} \lan\j(0)|\j(t)\ran$ is expressible  
as a line integral of what is called the connection ${A'}(s) = i \lan \j'(s)|\partial_{s} | \j'(s) \ran$:
\begin{align} \label{gint}
	\text{Arg} \lan \j(0) | \j (t) \ran =  \int_{0}^{\Lambda} ds \: {A'}(s). 
\end{align} 
This shows that the geometric phase (\ref{gpcont}) comprises of two line integrals of the connection ${A}(s)$: (a) along the time evolution curve defined by states $|\j(s)\ran$  connecting state $|\j(0)\ran \lan \j(0)|$ to $|\j(t)\ran \lan \j(t)|$ and (b) returning to $|\j(0)\ran \lan \j(0)|$ along the geodesic curve $|\j'(s)\ran \lan \j'(s)|$. So the expression (\ref{gpcont}) can now be written in a manifestly gauge invariant form as a closed line integral over the time evolution and geodesic curves:
\begin{align} \nonumber
	\gamma_{g}(t) &= i \int_{0}^{t} dl \: \lan \j(l)| \partial_{l} | \j(l) \ran + i\int_{0}^{\Lambda} ds \: \lan \j'(s)| \partial_{s} | \j'(s) \ran \\ &= \oint_{C} \: ds A(s).
\end{align}
It must be emphasised that while we are using states $|\j(l)\ran$ and $|\j'(s)\ran$ to express the geometric phase, owing to the local gauge invariance one learns that it actually depends only on the respective density matrices.   

The notion of geometric phase as summarised here can also be understood in a rigorous mathematical manner using the language of fibre bundles as 
summarised in the Appendix~B.

\begin{figure}
	\begin{center}
		\includegraphics[width=0.6\linewidth]{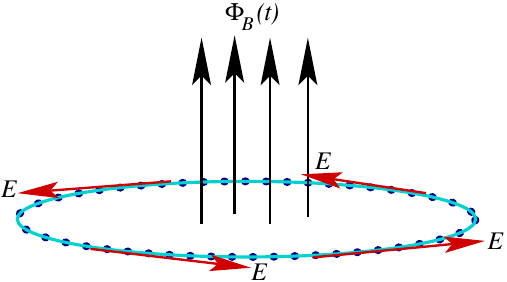}
		\caption{Schematic representation of the periodic lattice system studied in this paper. The cyan curve with blue solid circles represents the 
			periodic lattice, whereas the black arrows depict the magnetic flux $\Phi_{B}(t)$. The red arrows show the tangential electric field experienced by the electrons in the lattice.}
		\label{figlattice}
	\end{center}
\end{figure}

\section{Periodic potential problem \label{gpsec}}

Consider a (spinless) charged particle of mass $\m$ and charge $e$ in 1D under the influence of a periodic potential $V(x)$ with lattice constant $a$. It is assumed that the periodic potential arises due to the ions, and we shall be working in the rest frame of ions. We are assuming 
a periodic boundary condition (PBC), so that the system can be thought of as forming a ring. We allow a linearly time-varying magnetic flux $\Phi_{B}(t)$ to pierce the ring, while giving rise to a weak tangential electric field $E$ (see Fig. \ref{figlattice}). The particle dynamics in such a system is described by the Hamiltonian: 
\begin{align} \label{hamal}
	\hat{H}_{\a(t)} = \frac{1}{2 \m} \left(\hat{p} + \hbar \a(t)\right)^2 + V(\hat{x}),
\end{align}  
where the time-dependent vector potential $A(t) = -Et$ and $\a(t) = - eA(t)/\hbar$. 

The problem of particle motion in a periodic potential in the presence of a uniform electric field $E$ is a well studied one \cite{ashcroftbook, grecchi2001, schonhammer2001,hartmann2004}. In the literature, such a system is usually studied using the time independent Hamiltonian:
\begin{align} \label{dipolH}
	\hat{H}_{\varphi} = \frac{1}{2 \m} \hat{p}^2 + V(\hat{x}) - e E \hat{x},
\end{align}
wherein one works in the gauge $A=0$ with scalar potential $\varphi = - E {x}$. Clearly this Hamiltonian does not respect the periodicity of the lattice potential, since the spatial translation operator over a unit cell $\hat{T}_{x}(a) = e^{i \frac{\hat{p}}{\hbar}a}$ does not commute with $\hat{H}_{\varphi}$. This fact leads to the well known $k$-acceleration theorem \cite{ashcroftbook,schonhammer2001, grecchi2001} in such a gauge.

Motivated by Zak \cite{zak1989}, here we work with the 
gauge $A(t) = -Et$ and scalar potential $\varphi = 0$, so that the 
periodicity of potential $V(x) = V(x+a)$ is respected. As we will see below,
the $k$-acceleration theorem takes on a more subtle aspect in this gauge. 
The Hamiltonian 
(\ref{hamal}), while commuting with $\hat{T}_{x}(a)$, admits normalized simultaneous instantaneous eigenstates $\Psi_{nk_{m}\a}$ which solve: 
\begin{align} \label{def1}
	\hat{H}_{\a} \J_{nk_{m}\a}(x) = E_{nk_{m}\a} \J_{nk_{m}\a}(x), \\
	\hat{T}_{x}(a) \J_{nk_{m}\a}(x) = e^{i k_{m} a} \J_{nk_{m}\a}(x), \label{def2}
\end{align}
where $n$ is the band index. Owing to PBC, we have $\J_{nk_{m}\a}(x + N a) = \J_{nk_{m}\a}(x)$, so that each band consists of exactly $N$ states with wave vector (quantum number) $k_{m} = \frac{2 \pi}{N a} m$, where $m = 0,1,\cdots,N-1$. From (\ref{def2}), we immediately see that states $\J_{nk_{m}\a}(x)$ and $\J_{nk_{m+N}\a}(x)$ have the same $\hat{T}_{x}(a)$ eigenvalue, which is $e^{i k_{m} a}$. This in turn dictates that the normalized states $\J_{nk_{m}\a}(x)$ and $\J_{nk_{m+N}\a}(x)$ must be linearly dependent, so that \cite{ashcroftbook}:
\begin{align}\label{chidef}
	\J_{nk_{m+N}\a}(x) = \J_{nk_{m}\a}(x) e^{i \chi}, 
\end{align}
where $\chi$ is some arbitrary real number. So the states $\J_{nk_{m+N}\a}(x)$ and $\J_{nk_{m}\a}(x)$ describe the same physical state, as the corresponding density matrices are identical. It is often assumed that $\chi = 0$, a choice of convention which is referred to as the \emph{periodic gauge} condition~\cite{dalibard,resta2000manifestations}. Clearly, all the physical observables must be insensitive to the value of the unphysical phase $\chi$, a requirement not obeyed by the Zak phase, as we shall soon see.

As noted above, by a judicious choice of gauge, which ensures that the spatial periodicity of the system is not spoiled by the electromagnetic field, we have $[ \hat{H}_{\a} , \hat{T}_{x}(a) ] = 0$. This straight away shows that $\hat{T}_{x}(a)$ is conserved quantity under time evolution. As a result, if the system is prepared initially in the state with wave vector $k_{l}$,  then it is forbidden to evolve into any other state with wave vector $k_{l'}$ in any band at any time $t$. This is a manifestation of the fact that the wave vector $k_{l}$ is a conserved quantum number under evolution, in the present gauge choice wherein $\varphi=0$. This is in sharp contrast with the well known result of $k$-acceleration theorem in the gauge $A=0$. It must be noted that the $k$-conservation law does not prohibit the system from evolving to the state $| \J_{n' k_{l}\a(t)} \ran$ from the initial state $| \J_{n k_{l}\a(0)} \ran$ in a different band $n'$ with energy $E_{n' k_{l}\a} \neq E_{n k_{l}\a}$. However, if the external field is sufficiently weak, then the evolution to other band states is energetically suppressed, and such a transition can be ignored in the leading order.

The Hamiltonian (\ref{hamal}) has very interesting property under time evolution. The vector potential at certain discrete times $t$ can be written as a gauge transformation:
\begin{align} \nonumber
	A(t) = A(0) + \frac{i \hbar}{e} U^{\dagger}(x,t) \partial_{x} U(x,t),
\end{align} 
where
\begin{align} \nonumber
	U(x,t) = \exp \left( \frac{i}{\hbar} eEtx\right),
\end{align}
and $A(0)=0$ by virtue of its definition. Under such a transformation, the momentum operator transforms as: $\hat{p} - e A(t) = U^{\dagger}(x,t) \hat{p}\:U(x,t)$, which allows 
the Hamiltonian at some time $t$ and at $t=0$ to be unitarily connected:
\begin{align}
	\hat{H}(t) = U^{\dagger}(x,t) \hat{H}(0) U(x,t).
\end{align}
The gauge transformation $U(x,t)$ must respect the  PBC: $U(x,t) = U(x+L,t)$ in order to be a well defined operator. It is evident that only for time $t= j \t$ ($j$ is an integer), is the PBC respected, where 
\begin{align}
	\t = \frac{2 \pi \hbar}{eE L}.
\end{align} 
This shows that the Hamiltonian $\hat{H}(j \t)$ (for different $j$s) are physically the same (they are \emph{gauge equivalent}), their spectra are identical. Moreover their instantaneous eigenstates are related to each other by the gauge transformation:
\begin{align} \label{statec}
	\J_{nk_{m}\a(j\t)} (x)&= U^{\dagger}(x,j \t) \J_{nk_{m+j}\a(0)}(x) \\
	&= \exp \left( - i \frac{2 \pi x j}{L}  \right) \J_{nk_{m+j}\a(0)}(x),
\end{align}   
as also the energies $E_{nk_{m}\a(j\t)} = E_{nk_{m+j}\a(0)}$.
The gauge transformation factor $U(x,j \t) = e^{i \frac{2 \pi x}{L}j}$ has a very interesting topological property. It is a function of $x$, 
albeit with the PBC, implying that the points $x=0$ and $x=L$ are identified since $U(0,j \t) = U(L,j \t)$. This shows that it lives on a circle with circumference $L$. Now, $U(x,j \t)$ by definition is a phase and takes values only on the unit circle in the complex plane. So $U(x,j \t)$ is a map from one circle (with circumference $L$) to the unit circle. Such maps are classified in terms of homotopy classes~\cite{nakahara2003}, with each of them characterized by an integer called the \emph{winding number}, which measures the number of times one circle is wound on another. This shows that the integer $j$ appearing in $U(x,j \t)$ is actually the winding number; under one rotation in $x$ space, the factor $e^{i \frac{2 \pi x }{L}j}$ completes $j$ rotations of the unit circle. As a result, it is not possible to continuously deform $U(x,j \t)$ to some $U(x,j' \t)$ for $j \neq j'$. The class of such gauge transformations, which can not be continuously deformed into the identity (gauge) transformation, is often referred to as \emph{large gauge transformations}. 

%The system owing to the PBC admits the existence of the large gauge transformations $U(x,j \t)$, which are topologically non-trivial in the sense that each $U(x,j \t)$ belongs to different homotopy class with the winding number $j$.   

\section{Geometric phase in the periodic potential problem \label{gppsec}}

\begin{figure}
	\begin{center}
		\includegraphics[width=0.6\linewidth]{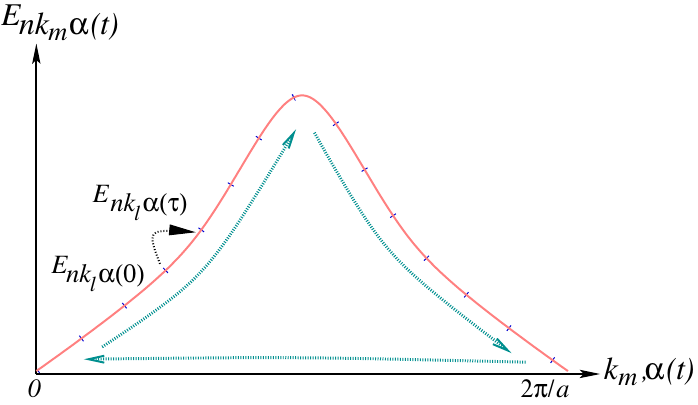}
		\caption{Schematic representation of the change in energy of the particle in the course of adiabatic evolution. The red curve represents the dispersion curve $E_{nk_{m}\a(t)}$ for some generic band $n$. The black arrow depicts the change in the energy over time $\t$. Whereas the blue arrows indicate that the particle returns to its initial energy after time $N \t$.}
		\label{figevol}
	\end{center}	
\end{figure}

As found earlier that if the system is initially prepared in the instantaneous eigenstate $| \Phi (0) \ran = | \J_{n k_{l} \a(0)} \ran$, then it is constrained to evolve with the same quantum number $k_{l}$ at any other time $t$. So the state of the system $| \Phi (t) \ran$ essentially evolves \emph{adiabatically}~\cite{messiah1999, tong2007} in the presence of a weak electromagnetic field, following the instantaneous eigenstate $| \J_{n k_{l} \a(t)} \ran$ along with an overall phase:
\begin{align} \label{tevol}
	| \Phi (t) \ran = e^{i \phi(t)} | \J_{n k_{l} \a(t)} \ran.
\end{align} 
The phase factor is given by $\phi(t) = i\int_{0}^{t} ds \: \lan \J_{n k_{l} \a(s)} | \frac{\d}{\d s} |\J_{n k_{l} \a(s)} \ran - \frac{1}{\hbar} \int_{0}^{t} ds \: E_{n k_{l} \a(s)}$~\cite{berry1984, messiah1999}.
In the light of (\ref{statec}), this takes the form:
\begin{align} \label{statet}
	| \Phi ( j \t) \ran = e^{i \phi(j \t)} \hat{U}^{\dagger}(x,j \t) |\J_{n k_{l+j}\a(0)} \ran,
\end{align}
which shows that the system, which was initially in the eigenstate with a wave vector $k_{l}$, in the course of adiabatic evolution goes into the eigenstate with a wave vector $k_{l + j}$ (in the same band) after time $t=j\t$, modulo a large gauge transformation with the winding number $-j$. The Fig.~\ref{figevol} graphically depicts the adiabatic evolution of the system due to $\alpha(t)$. This remarkable result leads to the evaluation of the geometric phase acquired by the system in two distinct cases: the single-particle case and the filled band many-particle case.   

\subsection{Single-particle case \label{gpz}}

Relation (\ref{statet}) dictates that after time $N\t$, the state of the system is:
\begin{align} \label{psint}
	| \Phi ( N \t) \ran  = e^{i \chi} e^{i \phi(N \t)} \hat{U}^{\dagger}(x,N \t) |\J_{n k_{l}\a(0)} \ran,
\end{align}
indicating that the system returns to the initial state with a large gauge transformation. It may be noted that in general, $| \lan \J_{n k_{l}\a(0)} | \hat{U}^{\dagger}(x,N \t) |\J_{n k_{l}\a(0)} \ran | \neq 1$ which indicates that 
the initial and final states are not colinear:
\begin{align}
	| \Phi ( N \t) \ran \neq e^{i \theta} | \Phi (0) \ran,
\end{align}
and the corresponding density matrices are not identical. Thus, strictly speaking the system does not return to its initial state after time $N\t$. However, owing to the gauge transformation factor $U^{\dagger}(x,N \t)$ it is straightforward to see that the average of any observable $\hat{F}(\hat{x}, \hat{p} - e A(t))$ returns after time $N\t$:
%\begin{widetext}
\begin{align} \nonumber
	\lan \J_{n k_{l}\a(0)} | \hat{F}(\hat{x}, \hat{p} - e A(0)) | \J_{n k_{l}\a(0)} \ran = 
	\lan \J_{n k_{l}\a(N\t)} | \hat{F}(\hat{x}, \hat{p} - e A(N\t)) | \J_{n k_{l}\a(N\t)} \ran.
\end{align} 	
%\end{widetext}
So the states $| \Phi (0) \ran$ and $| \Phi (N\t) \ran$ while being non-colinear, nevertheless represent the same physical state of the system, albeit expressed in different gauges. Thus, the time evolution of the system in this case is found to be \emph{adiabatic} and \emph{cyclic} kind. It must be mentioned that this notion of cyclicity generalizes the existing notion in the literature~\cite{samuel1988,mukunda1993} based on the requirement of returning of the density matrix.       

This treatment immediately shows that:
\begin{align}
	\lan \hat{x} (N\t) \ran_{\Phi} = \lan \hat{x} (0) \ran_{\Phi},
\end{align} 
showing that the center of mass of the wavepacket $| \Phi (t) \ran$ indeed performs \emph{Bloch oscillation} with time period $N\t$ \footnote{It must be noted that the notion of position operator $\hat{x}$ in a system with PBC is well defined only when the system size $Na \rightarrow \infty$.}. It is clear that such an oscillation phenomenon, consisting of cyclic acceleration and deceleration, would also be displayed by the average electric current $-\frac{e}{\mu} \lan \hat{p} + \hbar \a(t) \ran_{\Phi}$.  
The existence of Bloch oscillation and its time period $N\t = \frac{2 \pi \hbar}{e E a}$, are in exact agreement with the well known findings in the usual $A=0$, $\varphi \neq 0$ gauge \cite{grecchi2001,schonhammer2001,hartmann2004}.

The geometric phase gained by the system after such a cyclic adiabatic evolution then straight away follows from (\ref{tevol}) and (\ref{gpcont}) and it reads:
\begin{align}
	\gamma_{g}(n) = \text{Arg} \lan \J_{n k_{l} \a{(0)}} | \J_{n k_{l} \a(N\t)} \ran  +
	i \int_{0}^{N\t} dt \: \lan \J_{n k_{l} \a(t)}| {\partial_t} | \J_{n k_{l} \a(t)} \ran. 
\end{align}  
Recollect that the Bloch state $| \J_{n k_{l} \a(t)} \ran$ is not strictly periodic under spatial translation by a unit cell distance, but returns with a phase $e^{i k_{l}a}$. The above geometric phase expression simplifies significantly if we employ cell periodic Bloch state, defined as: 
\begin{align} \label{udef}
	|u_{n k_{l} \a}\ran = e^{-i k_{l} \hat{x}}| \J_{n k_{l} \a} \ran,
\end{align}
which is periodic under a unit cell translation. From (\ref{def1}), it follows that $|u_{n k_{l} \a}\ran$ solves the eigenvalue problem for the driven Hamiltonian $\hat{H}_{k_{l} + \a}$ so that: $\hat{H}_{k_{l} + \a} |u_{n k_{l} \a}\ran = E_{nk_{l}\a}|u_{n k_{l} \a}\ran$. When we redefine $q = k_{l}+\a$, we see that the Hamiltonian $\hat{H}_{k_{l} + \a} \equiv \hat{H}_{q}$ is a function of $q$ (following $k_l$-dependence of the undriven Hamiltonian $\hat{H}_{k_l}$. The above argument further dictates that both the energy $E_{nk_{l}\a} \equiv E_{nq}$ and the cell periodic Bloch state $|u_{n k_{l} \a}\ran \equiv |u_{n q}\ran$ of the driven Hamiltonian are also functions of $q$. 
From (\ref{chidef}), a crucial relation for these Bloch states follows:
\begin{align} \label{urel}
	| u_{n}(q + \frac{2 \pi}{a}) \ran = e^{i \chi} e^{- i \frac{2 \pi}{a} \hat{x}} | u_{n} (q) \ran.
\end{align}
We set $k_{l} = 0$ without loss of generality, and employ the reparameterisation invariance of the geometric phase, which enables us to express $\gamma_{g}(n)$ in terms of $| u_{n} (\a) \ran$ while treating $\a$ as a parameter. This leads us to the expression for \emph{Pancharatnam-Zak phase} $\gamma_{g}(n)$ which is one of the main results of this paper:
\begin{align} 
	\gamma_{g}(n) = \text{Arg} \lan u_{n} (0) | u_{n} ({2 \pi}/{a}) \ran + i \int_{0}^{\frac{2 \pi}{a}} d \a \: \lan u_{n} (\a) | {\partial_\a} | u_{n} (\a) \ran.\label{bandgp}
\end{align}
This geometric phase correctly and consistently characterizes the band. The Pancharatnam-Zak phase so obtained above is independent of the total number of cells $N$ in the system, as it should be, since it captures the curvature of the state space of the system, which is solely determined by the Hamiltonian. 

\begin{figure}
	\begin{center}
		\includegraphics[scale=0.5]{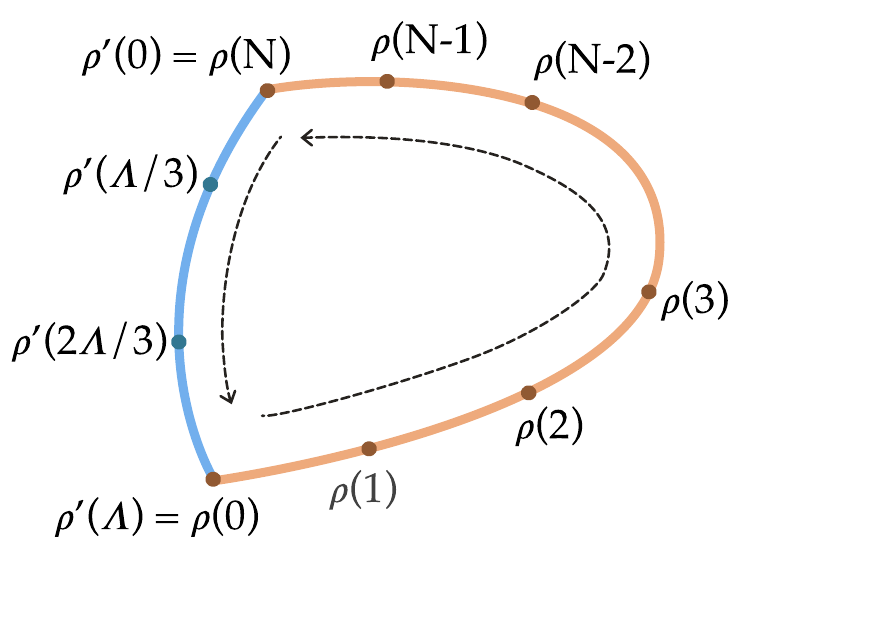}\caption{Schematic depiction of the curve along which the geometric phase integral in (\ref{closedfor}) is defined. Here density matrices $\rho(j)$ ($j=0,1,2,\cdots,N$) are the ones corresponding to states $|u_{n}(\frac{2 \pi}{Na}j)\ran$, specifying the adiabatic evolution (shown as brown curve). Whereas the density matrices $\rho'(l)$ correspond to the states $|u'_{n}(l)\ran$ which define the geodesic curve (shown as blue curve).\label{intfig}} 
	\end{center}	
\end{figure}

Invoking the treatment presented in section (\ref{panch}), one sees that the geometric phase (\ref{bandgp}) comprises of two line integrals of connection ${A}_{n}(s) = i \lan {u}_{n} (s) | \partial_{s} | {u}_{n} (s) \ran$ as shown in Fig. (\ref{intfig}): (a) along the adiabatic evolution curve $|u_{n}(q)\ran \lan u_{n}(q)|$ defined by states $|u_{n}(q)\ran$ which solve $\hat{H}_{q} |u_{n}(q)\ran = E_{nq}|u_{n}(q)\ran$ ($0 \leq q \leq \frac{2 \pi}{a}$) connecting state $|u_{n}(0)\ran \lan u_{n}(0)|$ to $|u_{n}(2 \pi/a)\ran \lan u_{n}(2\pi/a)|$, and (b) returning  to $|u_{n}(0)\ran \lan u_{n}(0)|$ via the geodesic curve $|u'_{n}(l)\ran \lan u'_{n}(l)|$ where $|u'_{n}(l)\ran = \frac{e^{i \theta l /\Lambda}}{\sin \Lambda} \left( \sin (\Lambda - l) |u_{n}(2\pi/a)\ran + e^{-i\theta} \sin (l) |u(0)\ran \right)$ (here $0 \leq l \leq \Lambda$ and $\theta = - \text{Arg}\lan u_{n}(0) | u_{n}(2 \pi/a)$). The Pancharatnam-Zak phase can thus be represented in a manifestly gauge invariant form as a closed line integral over adiabatic evolution and geodesic curves conjoint, to read:  
\begin{align} \label{closedfor}
	\gamma_{g}(n) = \oint_{C} dl \: {A}_{n}(l).
\end{align}

Owing to the fact that the state $| u_{n}(q) \ran$ and $e^{i \Lambda(q)} | u_{n}(q) \ran$, represent the same physical state of the system, since the corresponding density matrices are identical, one demands that a physically observable quantity must remain invariant under a gauge transformation $| u_{n}(q) \ran \rightarrow e^{i \Lambda(q)} | u_{n}(q) \ran$ for any choice of $\Lambda(q)$. It can be clearly seen from the above relation and (\ref{bandgp}) that the Pancharatnam-Zak phase is indeed insensitive to such a gauge transformation.

The Pancharatnam-Zak phase can be viewed as a 
cell periodic version of the Pancharatnam geometric phase (\ref{gpdef}) by defining:
\begin{align}
	\Delta_{N} = \lan u_{n,0} | u_{n,N} \ran \lan u_{n,N} | u_{n,N-1} \ran \cdots \lan u_{n,2} | u_{n,1} \ran \lan u_{n,1} | u_{n,0} \ran,
\end{align}   
where $| u_{n,i} \ran \equiv | u_{n}(\frac{2 \pi i }{N a}) \ran$. 
Evidently the cyclic nature of $\Delta_{N}$ ensures that
$\gamma_{g}(n)$ is invariant under local gauge
transformations $| u_{n,j} \ran \rightarrow e^{i \Lambda_{j}}| u_{n,j}
\ran$ (here $\Lambda_{j}$ are some arbitrary real numbers).
In the large 
$N$ limit, one immediately sees that expression (\ref{bandgp}) is indeed:
\begin{align} \label{pzdef}
	\gamma_{g}(n) = \lim_{N \rightarrow \infty} \text{Arg} \: \Delta_{N}.
\end{align}
It is also invariant under any unitary operation 
$\hat{U}$ (such that $\hat{U}^{-1} = \hat{U}^{\dagger}$) 
of the kind: $| u_{n,j} \ran \rightarrow \hat{U}| u_{n,j} \ran$, 
since such an operation preserves the value of all 
the amplitudes in $\Delta_{N}$. This crucially shows that the 
value of $\gamma_{g}(n)$ \emph{can not} be altered by 
changing the gauge convention and by translating the 
origin of the unit cell $| u_{n,i} \ran 
\rightarrow e^{-i \frac{\varepsilon}{\hbar} \hat{p}} | u_{n,i} \ran$ 
by distance $\varepsilon$. It is a geometric quantity that characterises the
band as a whole.

The spatial inversion (unitary) operator $\hat{\Pi}$ is defined such that $-\hat{x} = \hat{\Pi} \; \hat{x} \; \hat{\Pi}^{\dagger}$ and $- \hat{p} = \hat{\Pi} \; \hat{p} \; \hat{\Pi}^{\dagger}$. So for the lattices which are inversion symmetric, that is $V(-\hat{x}) = \hat{\Pi} \; V(\hat{x})  \hat{\Pi}^{\dagger} = V(\hat{x})$, one finds that $| {u}_{n}(-\k) \ran = \hat{\Pi} | {u}_{n}(\k) \ran$ which follows from the equation $\hat{H}_{\k} | {u}_{n}(\k) \ran = E_{n\k}| {u}_{n}(\k) \ran$. Using this in (\ref{bandgp}) along with the reparametrization invariance of $\gamma_{g}(n)$, one finds that the Pancharatnam-Zak phase for such a system is quantized:  
\begin{align}
	\gamma_{g}(n) = 0 \; \text{or} \: \: \pi.
\end{align}
This shows that the Pancharatnam-Zak phase in inversion symmetric lattices becomes a topological index, whose non-zero value corresponds to a topologically non-trivial band. 

The idea of evaluating the geometric phase gained by an electron in a 1D periodic lattice potential in the presence of a weak electromagnetic field has a long history starting from the celebrated work of Zak~\cite{zak1989}. The expression popularly referred to as the \emph{Zak phase}~\cite{vanderbilt2018,resta2000manifestations,niu,dalibard,moore2017} was obtained in this work, and it reads:
\begin{align} \label{zakp}
	\gamma_{Z}(n) = i \int_{0}^{\frac{2 \pi}{a}} d k \: \lan u_{n} (k) | \frac{\partial}{\partial k} | u_{n} (k) \ran,
\end{align}
wherein the states $| u_{n}(k) \ran$ are required to obey the \emph{periodic gauge condition}~\cite{vanderbilt2018,zak1989,resta2000manifestations,niu} whereby $\chi = 0$. 

It can be readily seen that the Zak phase (\ref{zakp}) is \emph{not} a gauge invariant object, and its value alters under a gauge transformation $| u_{n}(q) \ran \rightarrow e^{i \Lambda(q)} | u_{n}(q) \ran$, for any general $\Lambda(q)$. Furthermore even in the periodic gauge, Zak's expression (\ref{zakp}) does not yields the correct value for the geometric phase, since (\ref{zakp}) is devoid of the non-trivial contribution due to $\text{Arg} \lan u_{n} (0) | u_{n} ({2 \pi}/{a}) \ran$, which is captured by (\ref{bandgp}). 

The same conclusion can also be reached using a different representation of the Zak phase. As was shown by Resta~\cite{resta2000manifestations} and recently by Vanderbilt \cite{vanderbilt2018}, that the Zak phase (\ref{zakp}) can be written as argument of a \emph{non-cyclic object}:
\begin{align} \label{zakdef}
	\gamma_{Z}(n) = \lim_{N \rightarrow \infty} \text{Arg} \left( \lan u_{n,N} | u_{n,N-1} \ran \cdots \lan u_{n,2} | u_{n,1} \ran \lan u_{n,1} | u_{n,0} \ran \right) \biggl\lvert_{\chi=0},
\end{align} 
where $| u_{n,i} \ran \equiv | u_{n}(\frac{2 \pi i }{N a}) \ran$ and $| u_{n,N} \ran = e^{i \chi} e^{-i \frac{2 \pi}{a} \hat{x}} | u_{n,0} \ran$. It can be readily seen that this expression readily reproduces (\ref{zakp}), in $N \rightarrow \infty$ limit. This representation again shows that the Zak phase is a \emph{gauge dependent} construct, and its value depends on the choice of the unphysical quantity $\chi$. This feature is very uncharacteristic of a geometric object as also of a physical observable, which is required to be independent of the choice of gauge. 

The operation of shifting the origin of the unit cell by $\varepsilon$ distance is implemented by the transformation $| u_{n,j} \ran \rightarrow | u'_{n,j} \ran = e^{-i \frac{\varepsilon}{\hbar} \hat{p}} | u_{n,j} \ran$ (for $j = 0,1,2,\cdots,N-1$). The Zak phase defined using the transformed states is given by:
\begin{align}
	\gamma'_{Z}(n) = \lim_{N \rightarrow \infty} \text{Arg} \left( \lan u'_{n,N} | u'_{n,N-1} \ran \cdots \lan u'_{n,2} | u'_{n,1} \ran \lan u'_{n,1} | u'_{n,0} \ran \right) \biggl\lvert_{\chi=0}.
\end{align}
In light of relation (\ref{urel}), the above expression reads:
\begin{align}
	\gamma'_{Z}(n) &= \lim_{N \rightarrow \infty} \text{Arg} \left( \lan u'_{n,0} |e^{i \frac{2 \pi}{a} \hat{x}}| u'_{n,N-1} \ran \cdots \lan u'_{n,2} | u'_{n,1} \ran \lan u'_{n,1} | u'_{n,0} \ran \right) \\
	&= \lim_{N \rightarrow \infty} \text{Arg} \left(e^{i \frac{2 \pi}{a}\varepsilon} \lan u_{n,0} |e^{i \frac{2 \pi}{a} \hat{x}}| u_{n,N-1} \ran \cdots \lan u_{n,2} | u_{n,1} \ran \lan u_{n,1} | u_{n,0} \ran \right). 
\end{align}
The last expression clearly shows that under spatial translation of the unit cell by $\varepsilon$ distance, the value of the Zak phase indeed gets altered as:
\begin{align}
	\gamma_{Z}(n) \rightarrow \gamma'_{Z}(n) = \gamma_{Z}(n) + \frac{2 \pi}{a} \varepsilon,
\end{align}    
which is a well known result \cite{vanderbilt2018,zak1989,moore2017}.
So by a suitable choice of the origin of the unit cell, one can make the Zak phase attain any desired value. These observations ultimately overturn the assertion that the Zak phase is a geometric phase and a physical observable. The fact that the Zak phase depends on the gauge choice and the choice of the origin of the unit cell is well acknowledged in the literature ~\cite{vanderbilt2018,resta2000manifestations, niu, dalibard, atala2013, moore2017}.

In order to appreciate the construction of the Pancharatnam-Zak phase, it is instructive to consider its  behaviour under spatial translation of the unit cell $| u_{n,j} \ran \rightarrow | u'_{n,j} \ran = e^{-i \frac{\varepsilon}{\hbar} \hat{p}} | u_{n,j} \ran$ (for $j = 0,1,2,\cdots,N-1$). In terms of the transformed states, the Pancharatnam-Zak phase is given by:
\begin{align}
	\gamma'_{g}(n) = \lim_{N \rightarrow \infty} \text{Arg} \left( \lan u'_{n,0} | u'_{n,N} \ran \lan u'_{n,N} | u'_{n,N-1} \ran \cdots \lan u'_{n,2} | u'_{n,1} \ran \lan u'_{n,1} | u'_{n,0} \ran \right).
\end{align}
Employing relation (\ref{urel}), the above expression reads:
\begin{align}
	\gamma'_{g}(n) = \lim_{N \rightarrow \infty} \text{Arg} \left( \lan u'_{n,0} | e^{i \chi} e^{-i \frac{2 \pi}{a} \hat{x}}|u'_{n,0} \ran \lan u'_{n,0}| e^{-i \chi} e^{i \frac{2 \pi}{a} \hat{x}} | u'_{n,M-1} \ran \cdots \lan u'_{n,2} | u'_{n,1} \ran \lan u'_{n,1} | u'_{n,0} \ran \right).
\end{align} 
The structure of the first and second amplitudes in the above expression clearly shows that $\gamma'_{g}(n)$ is insensitive to the value of $\chi$. Invoking the relation $| u'_{n,j} \ran = e^{-i \frac{\varepsilon}{\hbar} \hat{p}} | u_{n,j} \ran$, one immediately sees that the $\varepsilon$ dependent phase factors from the first and second amplitudes precisely cancel each other, so as to yield:
\begin{align}
	\gamma'_{g}(n) = \gamma_{g}(n).	
\end{align}
This is a clear demonstration of the gauge invariance of the Pancharatnam-Zak phase and its invariance under spatial translation of the unit cell.

A careful observation shows that the expression (\ref{bandgp}) of the Pancharatnam-Zak phase coincides with the Zak phase (\ref{zakp}), \emph{not} in the periodic gauge $\chi = 0$, 
but in the gauge wherein $\chi$ is defined such that $\text{Arg} \lan u_{n} (0) | u_{n} ({2 \pi}/{a}) \ran = 0$. This gauge provides a way of using Zak's formula to yield the correct value of the geometric phase, even though its original derivation was done within the framework of periodic gauge wherein $\text{Arg} \lan u_{n} (0) | u_{n} ({2 \pi}/{a}) \ran \neq 0$ \cite{zak1989,vanderbilt2018}.

\subsection{Many-particle case}

The equation (\ref{statet}) states that if the system is prepared in the initial state $|\J_{n k_{l}\a(0)}\ran$, then after time $\t$ it adiabatically evolves as:  
\begin{align} \label{statett}
	| \Phi ( \t) \ran= e^{i \phi( \t)} \hat{U}^{\dagger}(x, \t) |\J_{n k_{l+1}\a(0)} \ran,
\end{align}
and the Hamiltonian returns modulo a large gauge transformation: $\hat{H}(\t) = \hat{U}^{\dagger}(x,\t) \hat{H}(0) \hat{U}(x,\t)$. This observation motivates one to consider the $N$-particle generalization of this problem, the case wherein the $n^{th}$ band is completely filled by $N$ non-interacting spinless fermions. In the literature, the single-particle case considered in the earlier section has attracted significant interest. The discussion of the many-particle case is essential for practical topological materials, such as 1D topological insulators and superconductors. It is the many-particle systems (e.g., filled bands of topological insulators) whose geometric and topological features are probed experimentally. In such many-particle systems, the band-gap between the bands of topological insulators or the pairing gap in topological superconductors plays a vital role for the topological protection and the validity of adiabatic conditions in the definition of geometric phase \cite{bernevig2013topological, asboth2016short}.

Let us consider the many-particle wavefunction $\bar{\Phi}_{n}$ representing such a filled band at any time $t$ in the adiabatic approximation, given by the Slater determinant:
%\begin{widetext}
\begin{align}\label{slater}
	\bar{\Phi}_{n}(x_{1},x_{2}, \cdots, x_{N};\a(t)) = \frac{1}{\sqrt{N!}} \begin{vmatrix}
		\Phi_{n k_{0} \a(t)}(x_{1}) & \Phi_{n k_{1} \a(t)}(x_{1}) & \cdots & \Phi_{n k_{N-1} \a(t)}(x_{1}) \\
		\Phi_{n k_{0} \a(t)}(x_{2}) & \Phi_{n k_{1} \a(t)}(x_{2}) & \cdots & \Phi_{n k_{N-1} \a(t)}(x_{2}) \\
		\vdots & \vdots & & \vdots \\
		\Phi_{n k_{0} \a(t)}(x_{N}) & \Phi_{n k_{1} \a(t)}(x_{N}) & \cdots & \Phi_{n k_{N-1} \a(t)}(x_{N})
	\end{vmatrix}.
\end{align}
%\end{widetext}
Here, $\Phi_{n k_{l} \a(t)}(x_{i})$ represents the $i^{th}$ particle wave function adiabatically evolving as per (\ref{tevol}). From here it follows that the many-particle wavefunction at time $j \t$ can be straightforwardly written as:
%\begin{widetext}
\begin{align*}
	\bar{\Phi}_{n}(x_{1},x_{2}, \cdots, x_{N};\a(j \t))  = &\frac{e^{i \Gamma(j \t)}}{\sqrt{N!}} G(x_{1},x_{2}, \cdots, x_{N};j \t) \\ & \times \begin{vmatrix}
		\Psi_{n k_{j} \a(0)}(x_{1}) & \Psi_{n k_{j+1} \a(0)}(x_{1}) & \cdots & \Psi_{n k_{N+j} \a(0)}(x_{1}) \\
		\Psi_{n k_{j} \a(0)}(x_{2}) & \Psi_{n k_{j+1} \a(0)}(x_{2}) & \cdots & \Psi_{n k_{N+j} \a(0)}(x_{2}) \\
		\vdots & \vdots & & \vdots \\
		\Psi_{n k_{j} \a(0)}(x_{N}) & \Psi_{n k_{j+1} \a(0)}(x_{N}) & \cdots & \Psi_{n k_{N+j} \a(0)}(x_{N})
	\end{vmatrix}.
\end{align*}	
%\end{widetext}
The $N$-particle large gauge transformation $G$ is given by the product:
\begin{align}
	G(x_{1},x_{2}, \cdots, x_{N};\t) = \prod_{j=1}^{N} U^{\dagger}(x_{j},j \t),
\end{align}
whereas the phase factor $\Gamma(j \t)$ reads:
\begin{align} \label{manypp} 
	\Gamma(j \t) &= \sum_{l=0}^{N-1} \left( i \int_{k_{l}}^{k_{l+j}} d \a \: \lan u_{n} (\a) | {\partial_\a} | u_{n} (\a) \ran  - \frac{1}{\hbar}  \int_{0}^{j \t} dt \: E_{n k_{l} \a(t)} \right).
\end{align}
The identity $\Psi_{n k_{N+j} \a}(x_{i}) = e^{i \chi} \Psi_{n k_{j} \a}(x_{i})$ and the anti-symmetric nature of the Slater determinant yield:         
\begin{align} \nonumber
	\bar{\Phi}_{n}(x_{1},x_{2}, \cdots, x_{N};\a(j \t)) = & \: (-1)^{j(N-j)} e^{i j\chi} e^{i \Gamma(j\t)} \\& \times G(x_{1},x_{2}, \cdots, x_{N};j\t) \bar{\Phi}_{n}(x_{1},x_{2}, \cdots, x_{N};\a(0)).
\end{align}
Noting that the average of a single particle observable $\hat{F}(\hat{x}, \hat{p} - e A(\t))$ evolves as $\lan \J_{n k_{l}\a(\t)} | \hat{F}(\hat{x}, \hat{p} - e A(\t)) | \J_{n k_{l}\a(\t)} \ran = \lan \J_{n k_{l+1}\a(0)} | \hat{F}(\hat{x}, \hat{p} - e A(0)) | \J_{n k_{l+1}\a(0)} \ran$, one sees that the average \newline $\lan \bar{\Phi}_{n} (\a(t))| \hat{\mathscr{F}} | \bar{\Phi}_{n} (\a(t)) \ran$ of any $N$-particle observables $\hat{\mathscr{F}}$, for example the total Hamiltonian and momentum, return to itself after time $\t$. Generalizing the relation (\ref{gpcont}) for a filled band scenario, one finds that the geometric phase acquired by the band fermions evolving adiabatically till time $j\t$ reads:  
%\begin{widetext}
\begin{align} \label{gpMB}
	\Gamma_{g}(j\t) = \sum_{l=0}^{N-1} \left(\text{Arg} \lan u_{n} (k_{l}) | u_{n} (k_{l+j}) \ran + i \int_{k_{l}}^{k_{l+j}} d \a \: \lan u_{n} (\a) | {\partial_\a} | u_{n} (\a) \ran \right). 
\end{align}
%\end{widetext}
As noted above in this case the filled band system displays cyclicity even for time evolution $\t$. While each particle only traverses a segment of a closed curve
over the band, the collective state in equation (\ref{slater}) traces a closed curve 
to return back to its original state, resulting in a multi-particle geometric phase (\ref{gpMB}). So the non-zero geometric phase acquired by the filled band state in an evolution for time $\t$ results from an addition of the geometric phases acquired by each constituent single particle states. Evidently the phase acquired by the filled band state after evolution till time $N \t$ reads:    
\begin{align} 
	\Gamma_{g}(N \t) = N\gamma_{g}(n). 
\end{align}
This is an expected result since each of the fermions is evolving independently in this non-interacting system, giving rise to Pancharatnam-Zak phase $\gamma_{g}(n)$, which all add up to yield this result.  
The geometric phase for filled bands has been studied for some time  now~\cite{resta1994macroscopic,resta2000manifestations,resta1998}. We emphasize that the topological properties of the bands is characterized by 
$\gamma_{g}(n)$ or $\Gamma_{g}(N \t)$ per particle. %However, such a contribution of Fermi-Dirac statistics and the total particle number to the geometric phase has not been reported earlier.
Such many-particle geometric phase in optical systems has been studied earlier theoretically~\cite{mehta2010} and later confirmed experimentally~\cite{martin2012,satapathy2012} using intensity interferometry.

\section{Explicit example \label{eexam}} 
\begin{figure}
	\begin{center}
		\includegraphics[width=0.5\linewidth]{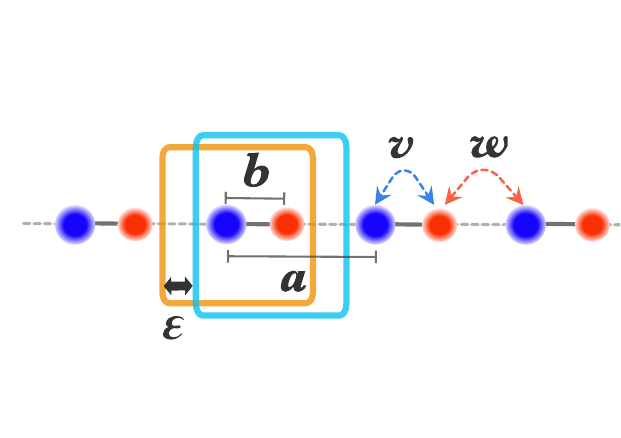}
		\caption{Schematic representation of Su-Schrieffer-Heeger (SSH) model with a lattice constant $a$ and a unit cell consisting of two different atoms (shown as red and blue disks) separated by a distance $b$. Here, the intracell and intercell hopping amplitude are $v$ and $w$, respectively. The yellow and blue rectangles represent two different choices of defining the unit cell, whose origins are separated by a distance $\varepsilon$.}
		\label{schlattice}
	\end{center}
\end{figure}

The SSH model is a 1D lattice of atoms with an unit cell consisting of two atoms, as depicted in Fig. \ref{schlattice}. This model is formulated within the tight-binding approximation with  nearest-neighbour couplings between the atoms~\cite{asboth2016short,dalibard}. In the recent years, there have been many experimental realizations of this model in various set-ups~\cite{atala2013, Blanco2016, Jean2017}. The Hamiltonian describing the model reads as:
\begin{align} \nonumber
	\hat{H}_{SSH} =  \sum_{m=-N/2}^{N/2}  \left(  - v |m a + r_{\a} \ran \lan m a + r_{\b} |  - w | (m+1) a + r_{\a} \ran \lan m a + r_{\b} | + \text{h.c.} \right). \label{sshh}
\end{align}   

Here, $r_{\a,\b}$ represent the coordinates of the two atoms respectively within the unit cell, whereas $a$ is the distance between the unit cells. The spatially localized electron state on atom in $m^{\mathrm{th}}$ unit cell at site $r_{\a,\b}$ is described by $|m a + r_{\a,\b} \ran$, whereas the distance between the two atoms in the unit cell is $b=r_{\b} - r_{\a}$. The parameter $v$ is the intracell electron hopping amplitude, whereas $w$ is the intercell hopping amplitude. The system consists of $N$ cells with PBC. Going over to the momentum space allows one to define the free particle states $| k \ran_{\a,\b}$ as:
\begin{align}
	| k \ran_{\a,\b} = \frac{1}{\sqrt{N}} \sum_{m=-N/2}^{N/2} \: e^{i k (m a + r_{\a,\b})} |m a + r_{\a,\b} \ran, 
\end{align}          
so that the above Hamiltonian reads:
\begin{align}
	\hat{H}_{SSH} = \sum_{k \in FBZ} \left( \begin{matrix} | k \ran_{\a}, & | k \ran_{\b} \end{matrix} \right) \left[ \mathscr{H}(k) \right] \left( \begin{matrix} _{\a}\lan k |  \\ _{\b}\lan k | \end{matrix} \right).
\end{align}
Here, the sum is over all the allowed values of $k$ in the first Brillouin zone (FBZ). The $2 \times 2$ matrix $\left[ \mathscr{H}(k) \right]$ has only two non-vanishing off diagonal elements $\mathscr{H}_{\a \b} = \mathscr{H}^{\ast}_{\b \a} = x(k) + i y(k) = \left(- v e^{i k b} - w e^{i k (b-a)} \right)$. This can be diagonalized to find two eigenvalues $E_{\pm}(k) = \pm E(k)$, where $E(k) = \sqrt{x^{2}+ y^{2}} = \sqrt{v^{2} + w^{2} + 2 v w \cos (k a)}$. The corresponding eigenvectors $[u_{\pm}(k)]$ are given by:
\begin{align}
	[u_{\pm}(k)] = \left( \begin{matrix}
		u_{1,\pm}(k) \\ u_{2,\pm}(k)\end{matrix} \right) = \frac{1}{\sqrt{2}} \left( \begin{matrix}
		\pm \frac{(x(k) + i y(k))}{E(k)} \\ 1
	\end{matrix} \right). 
\end{align} 
Note that, there is an ambiguity (upto a local gauge transformation) in defining these eigenvectors, since $[u_{\pm}(k)]$ and $e^{i \theta(k)}[u_{\pm}(k)]$ (where $\theta(k)$ is any general function of $k$) both solve the eigenvalue problem for $[\mathscr{H}(k)]$ for the same eigenvalues. It follows that the Hamiltonian (\ref{sshh}) diagonalises in terms of $| \Psi_{\pm}(k) \ran$:
\begin{align}
	\hat{H}_{SSH} = \sum_{k \in FBZ} \:  E_{\pm}(k) | \Psi_{\pm}(k) \ran \lan \Psi_{\pm}(k) |,
\end{align} 
which are defined as:
\begin{align} \nonumber
	| \Psi_{\pm}(k) \ran = u_{1,\pm}(k) | k \ran_{\a} + u_{2,\pm}(k) | k \ran_{\b}. 
\end{align}
This allows one to determine the cell periodic Bloch states $| u_{\pm}(\k) \ran$ as:
\begin{align}
	| u_{\pm}(\k) \ran = u_{1,\pm}(\k) | 0 \ran_{\a} + u_{2,\pm}(\k) | 0 \ran_{\b},
\end{align}
so that the Pancharatnam-Zak phase 
$\gamma_{g}(\pm)$ from (\ref{bandgp}) is given by:
%\begin{widetext}
\begin{align} \label{gpu}
	\gamma_{g}(\pm) = \text{Arg}\left( u^{\ast}_{1,\pm}(0) u_{1,\pm}(2\pi/a) + u^{\ast}_{2,\pm}(0) u_{2,\pm}(2\pi/a) \right) + i \int_{0}^{2\pi/a} d \k \: 
	\left( u^{\ast}_{1,\pm} {\d}_{\k} u_{1,\pm} + u^{\ast}_{2,\pm} {\d}_{\k} u_{2,\pm} \right).
\end{align}	
%\end{widetext}
Evidently, the ambiguity of local phase factor $e^{i \theta(\k)}$ in the definition of $| u_{\pm}(\k) \ran$ does not affect $\gamma_{g}(\pm)$. It can be readily checked that the contributions from the first and second terms in the expression (\ref{gpu}) arising from such a phase factor get exactly cancelled, displaying yet again the gauge invariance of the Pancharatnam-Zak phase. It can be immediately seen that the above expression is also invariant under spatial translation operation.  

This is to be contrasted with the range of values of the Zak phase $\gamma_{Z}(\pm)$ reported in literature~\cite{asboth2016short,dalibard,atala2013,niu} which is arising due to the different choices in defining $[u_{\pm}(\k)]$ and the choice of the origin of the unit cell. 

From relation (\ref{gpu}), we find that $\gamma_{g}(\pm)$ takes two values: (a) it is equal to $\pi$ when $v/w < 1$, and (b) it is equal to $0$ when $v/w > 1$; when $b < a/2$. Interestingly, when $b > a/2$, the system resembles itself with $b < a/2$ case, albeit with the roles of $v$ and $w$ now interchanged. Thus, one finds that (a) $\gamma_{g}(\pm) = \pi$ when $v/w > 1$, and (b) $\gamma_{g}(\pm) = 0$ when $v/w < 1$ for  $b > a/2$. 

This discussion shows that the two discrete phases in the 
SSH model are properly captured by the geometric phase $\gamma_{g}$, the 
topologically trivial phase exists when $\gamma_{g}(\pm)=0$, and 
the topologically nontrivial phase exists when 
$\gamma_{g}(\pm) = \pi$. Interestingly, we find that the 
(non)trivial value of $\gamma_{g}(\pm)$ also correctly 
identifies the (presence)absence of the 
gapless edge states in the SSH model, when defined with 
open boundary condition \cite{asboth2016short}.

It is worthwhile to consider the work of Atala \emph{et al.}, wherein the experimental observation of the Zak phase in this model was reported~\cite{atala2013}. The SSH model in their experimental setup was realized in an optical lattice setup, and the difference in the geometric phase 
between the topologically trivial  and non-trivial phase of the 
system was observed to be equal to $\pi$. The variability of the actual value of the Zak phase $\gamma_{Z}(\pm)$ in the two phases of the model, due to dependence on gauge and on the unit cell parametrization was acknowledged in this work~\cite{asboth2016short,niu,dalibard,atala2013}. 
However it was also noted that the difference of the Zak phase between the 
topologically trivial  and non-trivial phases of the 
system is equal to $\pi$, and is immune to such 
gauge and unit cell parametrization dependence. As a result 
the experimental measurement of the difference in the geometric phase was attributed to the observation of the difference in the Zak phase. Since the difference in the Pancharatnam-Zak phase between the 
topologically trivial  and non-trivial phases of the model is also 
equal to $\pi$, the experimental measurement of Atala \emph{et al.} is 
inconclusive in determining whether Zak phase or Pancharatnam-Zak phase 
is the correct expression for the geometric phase in such a system. These 
experiments are insensitive to the 
absolute value of the geometric phase.

\section{Discussion}
\label{dis}
In this paper, we provide a correct and consistent understanding of the notion of the geometric phase in 1D periodic lattice system weakly perturbed by electromagnetic field. The expression for the Pancharatnam-Zak phase, which is the geometric phase acquired by an electron traversing the Bloch band while moving in the lattice, is found, and the underlying mathematical structure is unveiled. Our work cures both the gross flaws of the popular Zak phase, its dependence upon the choice of origin of the unit cell and on the gauge choice. 

The Pancharatnam-Zak phase is a quintessentially geometric object insensitive to the choice of gauge and unit cell origin. In the case of systems with inversion symmetry, it is found that the Pancharatnam-Zak phase acts  as a topological index characterizing the band, can either be equal to $0$ or $\pi$. An explicit calculation of this geometric phase is demonstrated for the SSH model, and its absolute value is found to correctly predict the presence/absence of the gapless edge states. Our estimation this geometric phase is also found to be in agreement with the observation of Atala \emph{et al.}~\cite{atala2013} wherein the difference of the geometric phases in the topological and trivial phases of the model was measured. Nevertheless, it would be exciting to experimentally confirm the calculated values of the single-particle Pancharatnam-Zak phase separately in the topological and trivial phases of the SSH model by generalizing the current experimental schemes.

A many-particle generalization of this geometric phase for a filled band case is obtained, and its physical implications are highlighted. Our work would be useful in unambiguous characterization of many physical properties including quantized particle transport, edge modes and electrical polarization in 1D dielectric materials in terms of the geometric phase.

%remarkably it is found to be sensitive to the total number of the electrons as also its odd/even nature. Such contributions to the geometric phase are absent in the previous treatments dealing with filled band systems which are common in condensed matter physics~\cite{bernevig2013topological, asboth2016short}. It would also be interesting to observe such a geometric phase in the experimental set-up of~\textcite{atala2013} with multiple fermionic cold atoms in optical lattices. 

%This work provides a foundation for proper understanding of physical properties, like quantized particle transport, edge modes and electrical polarization in one-dimensional dielectric materials in terms of the geometric phase. It is straightforward to see that the Pancharatnam-Zak phase will play a central role in the modern understanding of electrical polarization, as pioneered by Resta~\cite{resta2000manifestations} and Vanderbilt~\cite{vanderbilt2018}.
%It is not difficult to foresee that Pancharatnam-Zak phase will correctly characterize the edge states in planar honeycomb lattice systems~\cite{delplace}, as also that of photonic waveguide lattices~\cite{longhi2013} and topological lasers \cite{Jean2017}. 

\section*{Acknowledgements}

Prof. Joseph Samuel was a part of this work in its early stages and has motivated us throughout the course of this work. DR gratefully acknowledges the funding from the Department of Science and Technology, India via the Ramanujan Fellowship. We thank Dr. Nikolai Sinitsyn for a critical reading of our manuscript.

%\disclaimer{Insert disclaimer text here.}
\appendix
\setcounter{figure}{0}
\renewcommand\thefigure{A\arabic{figure}}

\section*{Appendix A} \label{appa}

Let us consider a set of unit normalized states $|y(s)\ran$, which are continuously parametrized by a monotonically increasing real parameter $s$ ($0 \leq s \leq \Lambda$). Then the infinitesimal distance $dl$ between the states $|y(s)\ran$ and $|y(s+ds)\ran$ is defined as $dl^{2} = \lan \del y (s)| \del y (s)\ran$, where $|\del y(s) \ran = |y(s + ds) \ran - |y(s)\ran \simeq | \partial_{s} y(s)\ran ds$. This can be rewritten to the leading order to read:
\begin{align}
	dl^{2} = \lan \partial_{s} y(s)| \partial_{s} y(s) \ran ds^{2}.
\end{align}   
In quantum mechanics, the states $|y(s)\ran$ and $e^{i\Lambda(s)}|y(s)\ran$ (for some arbitrary function $\Lambda(s)$) represent the same quantum state of the system, and hence any physically meaningful quantity must be insensitive to such a gauge transformation. Unfortunately the metric $\lan \partial_{s} y(s)| \partial_{s} y(s) \ran$ is not invariant under such a gauge transformation. It can be salvaged with the replacement of partial derivative $\partial_{s}$ with that of the covariant derivative, $D_{s} = \partial_{s} - \lan y(s)|\partial_{s}|y(s)\ran$. The gauge covariant metric then obtained is $\lan D_{s} y(s)| D_{s} y(s) \ran$, so that the covariant distance $dl$ is now defined as:
\begin{align}
	dl^{2} = \lan D_{s} y(s)| D_{s} y(s) \ran ds^{2}.	
\end{align} 
This allows us to define a finite distance $L$ between the states $|y(0)\ran$ and $|y(\Lambda)\ran$, as one walks along the given curve $|y(s)\ran$:
\begin{align}
	L = \int_{0}^{\Lambda} ds \: \sqrt{\lan D_{s} y(s)| D_{s} y(s) \ran}.
\end{align}
It must be mentioned that this distance is invariant under the gauge transformation $|y(s)\ran \rightarrow e^{i \Lambda(s)}|y(s)\ran$ by design, and hence $L$ is actually the distance between the two density matrices $|y(0)\ran \lan y(0)|$ and $|y(\Lambda)\ran \lan y(\Lambda)|$, rather than the corresponding vectors. As shown in Refs.~\cite{mukunda1993, samuel1988}, functionally extremising this distance provides us with the unique shortest distance \emph{geodesic curve} $|\tilde{y}(l)\ran \lan \tilde{y}(l)|$, which is defined by the states $|\tilde{y}(l)\ran$ ($0 \leq l \leq \Lambda$):
\begin{align}
	|\tilde{y}(l)\ran = \frac{e^{i \theta l/\Lambda} }{\sin \Lambda} \left( \sin (\Lambda - l) |y(0)\ran  + e^{-i\theta} \sin (l) |y(\Lambda)\ran \right),
\end{align}
connecting initial state $|y(0)\ran$ to the final state $|y(\Lambda)\ran$. Here, the angle $\theta$ is the argument of the overlap $\theta = \text{Arg} \lan y(0)|y(\Lambda)\ran$. It must be noted that two curves $|\tilde{y}(s)\ran$ and $e^{i\Lambda(s)}|\tilde{y}(s)\ran$ represent the same geodesic curve $|\tilde{y}(l)\ran \lan \tilde{y}(l)|$ and the local gauge transformation factor can not change the distance $L$ traversed by them. These geodesic curves are of great interest in the context of geometric phase, since the phase $\theta = \text{Arg} \lan y(0)|y(\Lambda)\ran$ is expressible as a line integral along this geodesic curve:
\begin{align}
	\text{Arg} \lan y(\Lambda)|y(0)\ran = i \int_{0}^{\Lambda} dl \: \lan\tilde{y}(l)|\partial_{l}|\tilde{y}(l)\ran.
\end{align}
Note that in the above treatment, no restriction is assumed on the choice of the initial and final states, $|y(0)\ran$ and $|y(\Lambda)\ran$ respectively. Utilizing this freedom, one can choose any two states in the Hilbert space as the initial and final states, implying that the phase of the overlap between any two such states is indeed given by the line integral of what is called the connection ${A}(l) = i\lan{y}(l)|\partial_{l}|{y}(l)\ran$ evaluated along the geodesic connecting them.

\section*{Appendix B} \label{appb}
The purpose of this appendix is to clarify the mathematical structure behind the Pancharatnam-Zak phase and some remarks regarding the ambiguities of the Zak phase.

Consider some general quantum system at hand defined over the Hilbert space $\mathcal{H}$.  In the space of unit normalised states $ {\cal N}=\{|\psi \ran \in {\cal H}|\lan \psi|\psi \ran =1 \}$, we identify all the states which satisfy $|\psi'\ran = \alpha |\psi \ran$, where $\alpha$ is any complex number with unit modulus. This defines the projective Hilbert space, which is often called the Ray space ${\cal R}$. It is evident that this ray space, is in fact, the space of density matrices corresponding to each normalized state. So we now have a $U(1)$ fibre bundle ${\cal E}=\{{\cal N},{\cal R},\pi\}$, 
where $\pi: {\cal N}\rightarrow {\cal R}$ is the projection that takes each element  $ p \in{\cal N}$ to the fibre on which it lies. Vectors in the tangent space $T_p{\cal N}$ which project down to zero in ${\cal R}$ are called vertical vectors. This defines a natural connection on the fibre bundle: Horizontal subspaces at $p\in {\cal N}$ are defined as those orthogonal to vertical vectors at $p$. This connection is often called the \emph{universal connection} and has been
studied in Ref. \cite{bottchern}. It is called the universal connection since any $U(1)$ bundle over a manifold $B$ can be obtained by embedding $B$ in ${\cal R}$ and pulling back the structure on ${\cal E}$ \cite{narasimhan1961}.

Now given any smooth curve $c(t)$ ($0 \leq t \leq \Gamma$) in ${\cal R}$ and an initial point $p\in \pi^{-1}(c(0))$ in $\mathcal{N}$, we can define its unique lift say $\tilde{c}(t)$ to ${\cal N}$, given by vectors $|y(t)\ran$, such that $\lan y(t)|\partial_t|y(t)\ran = 0$. If $c(t)$ is a closed curve in ${\cal R}$, its lift $\tilde{c}(t)$ in general may be open, that is, the two ends of $\tilde{c}(t)$, specified by states say $|y(0)\ran$ and $|y(\Gamma)\ran$ are related by a unit modulus complex number $\delta$: $\lan y (0)| y(\Gamma)\ran = \delta$. This defines the \emph{holonomy} $\delta$ of the universal connection along the curve $c(t)$. 

The structure defined above is well studied in the mathematical literature. It is purely geometrical, and no reference to the dynamics has been made up till now. Invoking the discussion in section (\ref{panch}) let us identify the closed curve ${c}(t)$ defined by $|y(t)\ran \lan y(t)|$ with the closed curve formed by $|\j(t)\ran \lan \j(t)|$ as defined by the time evolution, conjoint with the geodesic $|\j'(t)\ran \lan \j'(t)|$. Then one immediately sees that the geometric phase (\ref{gpcont}) is intimately related to the holonomy $\gamma_{g} = \text{Arg} \: \delta$ of the universal connection.   

One can naturally apply this notion further without any ambiguities to the quantum system discussed in section (\ref{gppsec}) and identify the closed curve ${c}(t)$ with adiabatic evolution curve generated by cell periodic Bloch states $|u_{n}(\alpha) \ran \lan u_{n}(\alpha)|$  (for $0 \leq \alpha \leq 2\pi/a$) conjoint with the geodesic curve $|u'_{n}(l) \ran \lan u'_{n}(l)|$ ($0 \leq l \leq \Lambda$), as shown in Fig. (\ref{intfig}). This shows that the Pancharatnam-Zak phase as expressed in (\ref{closedfor}) measures the holonomy of the connection, and is indeed a mathematically well defined object. 

Moore \cite{moore2017comment} had argued that the notion of the geometric phase in the periodic lattice problem is intrinsically ambiguous. He explores a mathematical structure in which a Hilbert space is attached to \emph{each} $k$ point of the Brillouin zone (which is considered as the parameter space). This structure results in an ambiguity in the connection of
a bundle over the parameter space, which manifests as a coordinate and gauge dependence of the  geometric phase, coinciding with the findings of Zak \cite{zak1989}. The author made a case for the justification of these ambiguities on some physical grounds invoking the works of Resta \cite{resta2000manifestations}, and King-Smith \& Vanderbilt \cite{king1993theory}, on polarization. A moment's reflection will convince the reader that there cannot be any physical ground justifying the gauge dependence of an observable physical quantity like the geometric phase, which
has been measured in several experiments. 

From the treatment presented in this paper, it is amply clear that when one invokes the most general definition of the geometric phase, as summarised in section (\ref{panch}), no ambiguities of any kind are encountered, as we have explicitly shown.

\appendix

%\bibliographystyle{unsrt}
%\bibliography{gpref}

\end{document}